\documentclass[12pt]{article}
\usepackage[dvips]{graphicx}
\usepackage{color}
\usepackage{amsfonts,amssymb}
\begin{document}

\vspace*{1cm}
\begin{center}
{\Large \bf The NA49 Experiment at CERN: Spectral Analysis\\[1ex]
in the Unified Picture for Hadron Spectra}\\

\vspace{4mm}

{\large A. A. Arkhipov\footnote{e-mail: arkhipov@mx.ihep.su}\\
{\it State Research Center ``Institute for High Energy Physics" \\
 142281 Protvino, Moscow Region, Russia}}\\
\end{center}

\vspace{2mm}
\begin{abstract}
In this note we show how the experimental material given by NA49
Collaboration at CERN looks in the developed recently unified picture
for hadron spectra. It is found that the results of the experimental
studies by the NA49 Collaboration are excellently incorporated in the
unified picture for hadron spectra. Our analysis shows that
$\Xi$-like baryon states observed by the NA49 Collaboration are the
states living in the corresponding KK tower built in according to the
earlier established general, physical law.
\end{abstract}

\section*{}

In October of 2003 the NA49 Collaboration \cite{1} reported the
results of resonance searches in the $\Xi^{-} \pi^{-}$, $\Xi^{-}
\pi^{+}$, $\bar\Xi^{+}\pi^{-}$ and $\bar\Xi^{+}\pi^{+}$ invariant
mass spectra in proton-proton collisions at $\sqrt{s}=17.2$ GeV. An
evidence was found for the existence of a narrow $\Xi^{-}\pi^{-}$
baryon resonance with a mass of 1862 $\pm$ 2 MeV and a width below
the detector resolution of about 18 MeV. The statistical significance
was estimated to be 4 $\sigma$. This state, now denoted as
$\Phi^{--}$ by PDG, was interpreted as a candidate for the exotic
$\Xi_{3/2}^{--}$ baryon with strangeness  $S =-2$, isospin I =
$\frac{3}{2}$ and a quark content of ($dsds\bar{u}$). At the same
mass a peak was observed in the $\Xi^{-} \pi^{+}$ spectrum which was
interpreted as a candidate for the $\Xi_{3/2}^{\,0}$ member of the
isospin quartet with a quark content of ($dsus\bar{d}$). The similar
enhancements in the corresponding antibaryon spectra were also found
at the same invariant mass.

The events were recorded at the CERN SPS accelerator complex. The
interactions were produced with a beam of 158~GeV protons  on a
cylindrical liquid hydrogen target of  20 cm length and 2 cm
transverse diameter. The used data sample consists of about 6.5~M
events. The details of a reconstruction to find the $\Xi^{-}$
candidates are carefully described in original paper \cite{1}. To
search for the exotic $\Xi_{3/2}^{--}$ baryon the selected $\Xi^{-}$
candidates were combined with primary $\pi^{-}$ tracks. The resulting
$\Xi^{-}\pi^{-}$ invariant mass spectrum is presented in Fig.~1a
extracted from the original paper. The shaded histogram shows the
mixed-event background, obtained by combining the $\Xi^{-}$ and
$\pi^{-}$ from different events and normalizing to the number of real
combinations. As seen in Fig.~1a, a narrow peak above the background
is visible at approximately 1.86 GeV. It is pointed out that the mass
window 1.848 - 1.870 GeV contains 81 entries with a background of
about $B = 45$ events, the signal of $S = 36$ events has a
significance of 4.0 standard deviations calculated as $S/\sqrt{S+B}$.
This state was interpreted as a candidate for the $\Xi_{3/2}^{--}$
pentaquark.

The invariant mass distributions for $\Xi^{-}\pi^{+}$,
$\bar\Xi^{+}\pi^{-}$ and $\bar\Xi^{+}\pi^{+}$ are plotted in
Figs.~1b,c,d. As seen, the enhancements are evident in all three
spectra. Fits to the combined signal of the $\Xi_{3/2}^{--}$ and its
antiparticle and $\Xi_{3/2}^{\,0}$ and its antiparticle yielded peak
positions of 1862$\pm$2 MeV and 1864$\pm$5 MeV respectively.

The sum of the four invariant mass distributions from Fig.~1 is shown
in Fig.~2a. As pointed out, summing the four mass distributions
increases the statistical significance of the peak to 5.6 $\sigma$.
Fig.~2b shows the combinatorial background subtracted distribution. A
Gaussian fit to the peak yielded a mass value of 1862$\pm$2~MeV and a
FWHM = 17~MeV with an error of 3~MeV, largely due to the uncertainty
in the background subtraction. The systematic error on the absolute
mass scale was determined below 1~MeV.

In previous studies \cite{2} we have analyzed the recent results from
several experimental groups \cite{3} reported the first observations
of very narrow, manifestly exotic baryons, now denoted as the
$\Theta^+$ (Q=1, S=1) states, with the simplest quark assignment
($uudd\bar s$) decaying into $nK^+$ and $pK_S^0$. Our analysis was
performed in the framework of the unified picture for hadron spectra
developed early \cite{4}. It was found that all discovered $\Theta$
states were excellently incorporated in the unified picture for
hadron spectra. Moreover, it was still more non-trivial that a strong
correlation of the experimentally observed peaks in the mass spectra
with the calculated spectral lines have been seen in all discussed
experiments.

In this note we present the spectral analysis of the experimental
material given by NA49 Collaboration at CERN and show how it looks in
the unified picture for hadron spectra.

According to our theoretical concept \cite{4} we start with building
the Kaluza-Klein tower of KK-excitations for the $\Xi\pi$ system by
the formula

\begin{equation}\label{omegaJpsi}
M_n^{\Xi\pi} = \sqrt{m_{\Xi}^2+\frac{n^2}{R^2}} +
\sqrt{m_{\pi}^2+\frac{n^2}{R^2}}\,,\quad (n=1,2,3,...),
\end{equation}
where $R$ is the same fundamental scale established before (see
\cite{5} and references therein for the details), $\Xi=(\Xi^0,
\Xi^-,\bar\Xi^+)$, $\pi=(\pi^0, \pi^\pm)$, and $m_{\Xi^0}$ = 1314.82
MeV, $m_{\Xi^-}=m_{\bar\Xi^+}$ = 1321.34 MeV, $m_{\pi^0}$ = 134.9766
MeV, $m_{\pi^\pm}$ = 139.57018 MeV have been taken from PDG. The such
built Kaluza-Klein tower is shown in Table 1.

The spectral lines corresponding to KK excitations in the $\Xi\pi$
system taken from 4th column in Table 1 have been plotted in Fig. 3.
As is seen, the experimentally observed peak almost coincided with
the spectral line corresponding to the
$M_{11}^{\Xi\pi}$(1968-1875)-storey in KK tower for the $\Xi\pi$
system. What is more important, a strong correlation of the spectral
lines with the other peaks on the histogram is also clear seen in
Fig.~3. It should be emphasized that earlier we have already found
out that strong correlation more than once in our previous studies
\cite{2,5}. In fact, that correlation displays the existence of the
resonances observed in other experiments. In particular, it should be
noted that the peak of the earlier observed $\Xi(1530)$ state, which
is also visible in the NA49 Experiment, exactly coincided with the
spectral line corresponding to the $M_{4}^{\Xi\pi}$(1539-1548)-storey
in KK tower for the $\Xi\pi$ system.

Our conservative estimate for the widths of KK excitations looks like
\begin{equation}\label{width}
\Gamma_n \sim \frac{\alpha}{2}\cdot\frac{n}{R}\sim 0.4\cdot n\,
\mbox{MeV},
\end{equation}
where $n$ is the number of KK excitation, and $\alpha \sim 0.02$,
$R^{-1}=41.48\,\mbox{MeV}$ are known from our previous studies
\cite{4}. This gives $\Gamma_{11}(\Phi^{--}\rightarrow\Xi\pi)\sim
4.4$ MeV which is in agreement with the experimental estimate.

In summary, the results of the experimental studies by the NA49
Collaboration at CERN are excellently incorporated in the unified
picture for hadron spectra developed early. Our analysis shows that
$\Xi$-like baryon states observed by the NA49 Collaboration are the
states living in the corresponding KK tower built in according to the
earlier established general, physical law. We expect that new
experiments will appear in the near future to confirm these exciting
measurements.

\newpage
\vspace*{2cm}
\begin{center}
Table 1. Kaluza-Klein tower of KK excitations for $\Xi\pi$ system\\
and experimental data.

\vspace{5mm}
\begin{tabular}{|c|c|c|c|c|c|}\hline
 n & $ M_n^{\Xi^0 \pi^0}MeV $ & $ M_n^{\Xi^0 \pi^{\pm}}MeV $ & $
 M_n^{\Xi^\pm \pi^0 }MeV $ & $ M_n^{\Xi^\pm \pi^\pm}MeV $ & $
 M_{exp}^{\Xi \pi}\,MeV $  \\
 \hline
1  & 1456.68 & 1461.08 & 1463.20 & 1467.59 &  \\
2  & 1475.87 & 1479.80 & 1482.38 & 1486.31 &  \\
3  & 1504.28 & 1507.69 & 1510.78 & 1514.18 &  \\
4  & 1539.14 & 1542.07 & 1545.61 & 1548.54 & $\Xi$(1530) \\
5  & 1578.54 & 1581.07 & 1584.98 & 1587.51 &  \\
6  & 1621.30 & 1623.52 & 1627.71 & 1629.93 & $\Xi$(1620) \\
7  & 1666.71 & 1668.67 & 1673.08 & 1675.04 & $\Xi$(1690) \\
8  & 1714.30 & 1716.06 & 1720.62 & 1722.38 &  \\
9  & 1763.78 & 1765.36 & 1770.05 & 1771.64 &  \\
10 & 1814.92 & 1816.37 & 1821.14 & 1822.58 & $\Xi$(1820) \\
11 & 1867.58 & 1868.91 & 1873.74 & 1875.07 & ${\color{red}\Xi_{3/2}^{--}}$(1862)\\
12 & 1921.64 & 1922.86 & 1927.74 & 1928.96 &  \\
13 & 1977.00 & 1978.13 & 1983.03 & 1984.17 & $\Xi$(1950) \\
14 & 2033.58 & 2034.63 & 2039.54 & 2040.60 & $\Xi$(2030) \\
15 & 2091.31 & 2092.30 & 2097.20 & 2098.19 & $\Xi$(2120) \\
16 & 2150.12 & 2151.05 & 2155.95 & 2156.88 &  \\
17 & 2209.97 & 2210.85 & 2215.72 & 2216.60 &  \\
18 & 2270.80 & 2271.63 & 2276.47 & 2277.30 & $\Xi$(2250) \\
19 & 2332.56 & 2333.35 & 2338.16 & 2338.95 &  \\
20 & 2395.21 & 2395.96 & 2400.73 & 2401.48 &  \\
21 & 2458.71 & 2459.42 & 2464.15 & 2464.86 &  \\
22 & 2523.00 & 2523.69 & 2528.36 & 2529.05 & $\Xi$(2500) \\
23 & 2588.07 & 2588.72 & 2593.35 & 2594.00 &  \\
24 & 2653.86 & 2654.49 & 2659.06 & 2659.69 &  \\
25 & 2720.35 & 2720.95 & 2725.47 & 2726.07 &  \\
26 & 2787.50 & 2788.08 & 2792.54 & 2793.12 &  \\
27 & 2855.27 & 2855.83 & 2860.24 & 2860.80 &  \\
28 & 2923.65 & 2924.19 & 2928.54 & 2929.08 &  \\
29 & 2992.60 & 2993.12 & 2997.41 & 2997.93 &  \\
30 & 3062.09 & 3062.59 & 3066.83 & 3067.33 &  \\ \hline
\end{tabular}

\end{center}

\newpage

\begin{figure}[htb]
\begin{center}
\includegraphics[width=\textwidth]{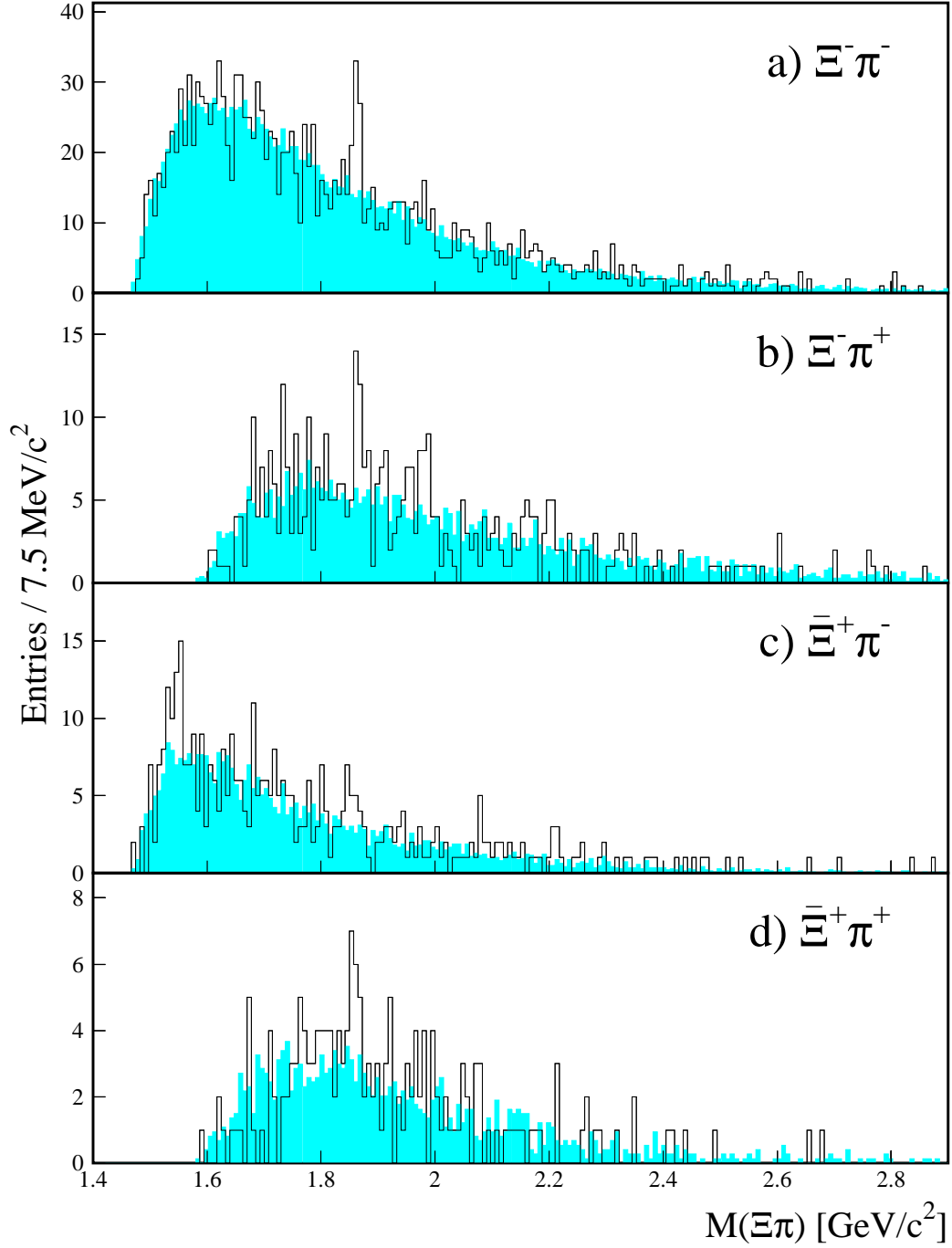}\label{fig1}
\end{center}
\caption{Invariant mass spectra presented in Ref. \cite{1} for
$\Xi^{-}\pi^{-}$ (a), $\Xi^{-}\pi^{+}$ (b),
$\overline{\Xi}^{+}\pi^{-}$ (c), and $\overline{\Xi}^{+}\pi^{+}$ (d).
The shaded histograms are the normalized mixed-event backgrounds. See
original paper \cite{1} for the details on selection cuts.}
\end{figure}

\newpage

\begin{figure}[htb]
\begin{center}
\includegraphics[width=\textwidth]{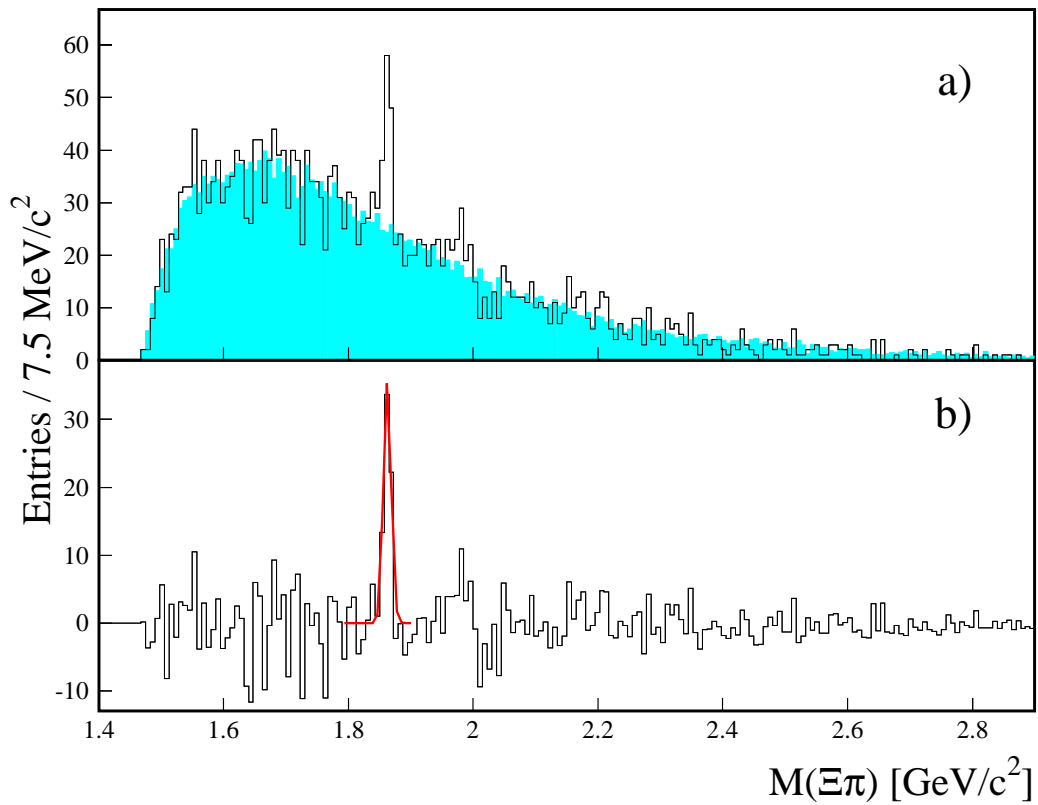}\label{fig2}
\end{center}
\caption{(a) The sum of the $\Xi^{-}\pi^{-}$, $\Xi^{-}\pi^{+}$,
$\overline{\Xi}^{+}\pi^{-}$ and $\overline{\Xi}^{+}\pi^{+}$ invariant
mass spectra presented in Ref. \cite{1}. The shaded histogram shows
the normalized mixed-event background. (b) Background subtracted
spectrum with the Gaussian fit to the peak. See original paper
\cite{1} for the details.}
\end{figure}

\newpage

\begin{figure}[htb]
\begin{center}
\includegraphics[width=\textwidth]{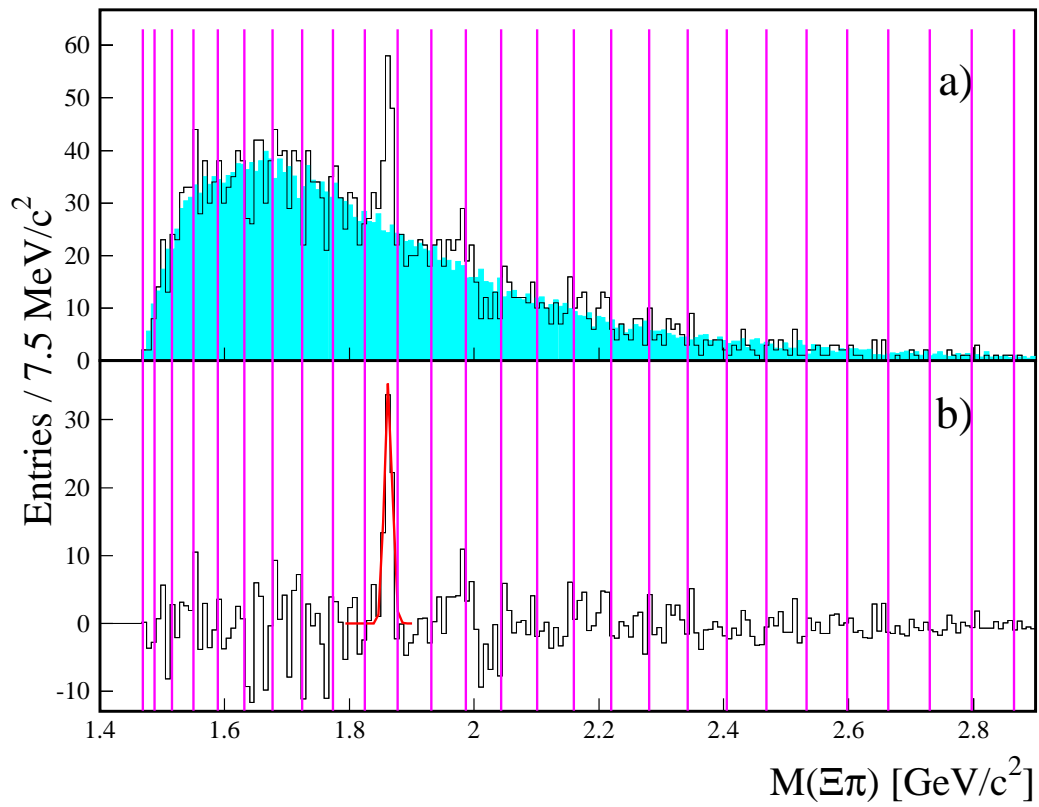}\label{fig3}
\end{center}
\caption{The same as in Fig.~2 but with the spectral lines
corresponding to 4th column in Table 1.}
\end{figure}

\end{document}